\documentstyle[prb,twocolumn,aps]{revtex}
\tighten
\begin{document} 
\title{A mesoscopic Tera-hertz pulse detector} 
\author{P. Orellana and F.Claro}
\address{$^{1}${\it  Departamento de F\'{\i}sica,
Universidad Cat\'olica del Norte , \\ 
Angamos 0610, Casilla 1280, Antofagasta , Chile} \\  
$^{2}${\it Departamento de F\'{\i}sica, Pontificia Universidad
Cat\'{o}lica de Chile,\\ 
Vicu\~{n}a Mackenna 4860, Casilla 306,
Santiago 22, Chile}}
\address{}
\address{\mbox{ }}
\address{\parbox{14cm}{\rm \mbox{ }\mbox{ }\mbox{ } 
We show that, under the passage of an electromagnetic terahertz pulse,
 an asymmetric double barrier device may act as an on/off current switch, 
depending on the bias. The time-dependent response of the device is discussed. 
}}
\address{\mbox{ }}
\address{\parbox{14cm}{\rm PACS numbers:  73.40.Gk,73.40.-c,72.15.Gd}}

\maketitle

\makeatletter 

\global\@specialpagefalse
\def\@oddhead{\underline{Accepted in Applied Physics Letters\hspace{281pt}
}}
\let\@evenhead\@oddhead
\makeatother

\narrowtext
 
It is well known that the characteristic frequencies of electronic processees  
in mesoscopic systems are in the terahertz range. This follows from transition  
energies that are typically in the meV scale. For example, instabilities in  
transport through asymmetric double barrier systems (ADBS) may give rise  
to terahertz oscillations. \cite{jensen,orellana} Other time dependent 
processes, such as  
the charging and discharging of the well in these systems, are also expected  
to take place in the range of picoseconds. This suggests that an ADBS may  
react as a  fast switch to the passage of terahertz pulse, a possibility that we  
explore in this work. 
 
ADBS are characterized by a bistable region of the bias, produced by charge  
accumulation in the space between the barriers (the well).\cite{gold} The collector barrier 
 is made wider with the purpose of increasing the lifetime of the resonance in  
the well, thus enhancing the amount of charge that is retained when current  
goes through.\cite{zas,orella} At a critical bias $V_{c\downarrow}$ the current drops abruptly due to a sudden  
emptying of the well, driven by an instability that may be understood using  
a nonlinear model. \cite{jensen} The dynamics is dominated by the accumulated  
charge which, in effect, lifts the bottom of the potential well thus retaining  
the resonance condition beyond $V_{c\downarrow }$ for ballistic transmission 
of an incoming electron. There is a point at which this charge is so  
large that it becomes favourable to spill it out, the well is emptied and the  
resonance condition is lost, followed by a large current drop. 
Similarly, when the well is uncharged it will remain so, even as the bias is  
decreased to values smaller than $V_{c\downarrow }$. At a second critical value  
$V_{c\uparrow}$ the resonance condition is fullfiled again and 
current now flows. This completes the bistable cycle,
a signature of which is the fact that $V_{c\downarrow}< V_{c\uparrow }$.

In this work we
propose that an ADBS  
device biased slightly below $V_{c\downarrow}$ will still undergo a transition,  
triggered by the passage of radiation in the terahertz region. 
The external field introduces
an additional oscillating field that may in effect bring the bias
to criticality. We also contend that
a radiation field may trigger the onset of resonant transport
if the system is biased slightly above $V_{c\uparrow}$. 
Optical radiation will have no effect in either case since the 
field then oscillates so  
fast that the electrons have no time to respond. Only at terahertz and lower  
frequencies would one expect the system to switch from a state of high  
current to one of low flow of electrons in the presence of an incident  
radiation pulse, or vice-versa.\cite{sollner,sollner1,chitta} 
 
Consider an ADBS under bias and in the presence of an electromagnetic 
field polarized along the z-axis, the growth direction. In order to study the time evolution 
of the device we adopt a first-neighbors tight-binding model for the hamiltonian. 
The radiation field enters as a space and time 
dependent voltage. To a good approximation the longitudinal degrees of 
freedom are decoupled from the transverse motion and may be treated 
independently. The probability amplitude $b_{j}^{\alpha}$ for an electron in 
a time dependent state $|\alpha>$, to be at plane $j$ along $z$, is determined 
by the equation of motion \cite{orellana} 
 
\begin{eqnarray} 
i\hbar \frac{db_{j}^{\alpha}}{dt}&=&(\epsilon_{j}(t)+ 
U\sum_{\beta}|b_{j}^{\beta}|^{2})b_{j}^{\alpha} \\ 
&+&v(b_{j-1}^{\alpha}+b_{j+1}^{\alpha}-2b_{j}^{\alpha}).  \nonumber 
\label{eq:time-eq} 
\end{eqnarray} 
 
\noindent In this expression $\epsilon_{j}(t)$ includes the fixed band contour, 
the external radiation-induced voltage $\delta E sin 2\pi\nu t$ and the applied
dc bias, the latter represented by a term linear in the spatial coordinate $j$. 
The sum over $\beta$ covers all 
occupied electron states and $v$ is the hopping matrix 
element between nearest 
neighbor planes. In writing Eq. (1) we have 
adopted a Hartree model for the electron-electron interaction, keeping just 
the intra-atomic terms as measured by the effective 
coupling constant U. \cite{orellana} 
As we will show in what follows, this nonlinear term is of key importance 
in the behavior of the system. 
 
The time dependent Eq.(1) is solved using a half-implicit numerical 
method 
which is second-order accurate and unitary \cite{orellana,mains}.  
Boundary conditions must be specified at the left ($z=-L$) and right 
($z=L$) edges of the structure. The approach taken here assumes that 
the wave function at time $t$ is given outside the structure 
by \cite{orellana,mains,tien} 
 
\begin{equation} 
b_j^\alpha (t)=(Ie^{ik_\alpha z_j}+R_j(t)e^{-ik_\alpha 
z_j})e^{-i\epsilon 
^\alpha t/\hbar },\;\;z_j\le -L 
\end{equation} 
 
\begin{equation} 
b^{\alpha}_{j}(t)=T_{j}(t)e^{ik^{\prime}_{\alpha } 
z_j}e^{-i\epsilon^{\alpha}t/\hbar}e^{-i\delta E cos(2\pi\nu 
t)/h\nu},\;\; 
z_j \ge L. 
\end{equation} 
 
\noindent 
Here $k_\alpha$ and $k^\prime_\alpha= \sqrt{2m^{*}[\epsilon^{\alpha}-%
\epsilon_{L}]}/\hbar$ are the wavenumbers of the incoming and outgoing 
states, respectively, with $\epsilon^{\alpha}=-4 v\;sin^2(k_\alpha a/2)$ the 
energy of the incoming particle. To model the interaction with the particle reservoir 
outside the structure the incident amplitude $I$ is assumed to be a constant 
independent of the coordinates. The envelope function of the reflected and 
transmitted waves, $R_{j}$ and $T_{j}$, are allowed to vary with $j$, 
however. Since far from the barriers these quantities are a weak function of 
the coordinate $z_j$ we restrict ourselves to the linear corrections only. 
This approximation is appropriate provided the time step $\delta t$ does 
not exceed a certain limiting value. For the results presented here, a 
value of $\delta t=3\times 10^{-17}$ s was found sufficient to eliminate 
spurious reflections at the boundary while maintaining numerical 
stability up to 40$\times 10^{-12}$s. 
In our numerical procedure the coefficients obtained without 
electromagnetic field for a given
dc bias are used as initial condition when the THz radiation is turned on. 
With $b_{j}^{\alpha}(t)$ known, the time dependent current at site j is 
 obtained numerically from \cite{mains} 
 
\begin{equation} 
J_j(t) = \frac{e}{\hbar}\int_{0}^{k_{f}}Im\{b_{j}^{*\alpha}(b_{j+1}^{%
\alpha}-b_{j}^{\alpha})\}(k_f^2-k_{\alpha}^2)dk_{\alpha}, 
\end{equation} 
 
\noindent 
where $k_{f}=\sqrt{2m^{*}\epsilon_f}/ \hbar$, with $\epsilon_{f}$ the  Fermi 
energy. 
 
We next apply our model to an asymmetric GaAs/AlGaAs double barrier 
structure, with emitter and collector barrier thicknesess of 1.12 nm 
(2 sites) and 3.36 nm, (6 sites) respectively, and a well thickness of 
11.2 nm (20 sites). The second barrier is made wider than the first in 
order to enhance the trapping of charge in the well. For this geometry the 
first resonance at zero bias occurs at 30 meV. The conduction band offset 
is set at 300 meV. The buffer layers are uniformly doped up to 3 nm 
from either barrier, so as to give a neutralizing free carrier concentration 
of 2$\times$ 10$^{17}$ cm$^{-3}$ at the contacts. In equilibrium, the Fermi level 
lies 19.2 meV above the asymptotic conduction band edge, so that the 
zero bias resonance lies well above the Fermi sea. The contribution to
the potential due to 
the applied bias is taken into account throught a term linear in 
$j$, which is assumed to arise from fixed charges. 
The parameter values in Eq. (1) are set at $v$ = -2.16 eV and $U$ = 100 meV. 
The latter was chosen phenomenologically so as to fit the 
experimental J-V characteristic for a GaAs devices. \cite{zas} 
The sample has 400 sites and the 
normalization of the wave functions is chosen so that charge from the 
electrons filling up to the Fermi energy exactly cancels the positive 
charge 
at the contacts. \cite{orella} We solved Eq. (1) using the procedure 
described above, for an energy mesh appropriate to compute the integral 
in Eq.(4). Good convergence was found for a mesh of 100 points.  

Figure 1 shows the current-voltage characteristic in the absence (solid line)  
and presence (dashed line) of radiation of amplitude $\delta E$ = 10 meV
at $\nu =$ 1 THz. In the latter case we exhibit an average of the current over 
time. Note that at 
the chosen values of parameters $V_{c\downarrow}= $ 0.320 V  and $V_{c\uparrow}$ = 0.282 V.  
It is clear from the figure 
that the radiation field narrows down the region of bistability, in agreement  
with previous results by I\~narrea and Platero 
\cite{platero}. This effect is to be expected since the time dependent 
field added to the bias brings the system periodically to criticality 
when the applied dc bias has not reached this condition yet, thus triggering the 
charging or discharging of the well. 

In Fig. 2 we show the time evolution of the charge density at the 
center of the well for different frequencies of the radiation field at 
$V$ = 0.310 V (empty circle in Fig. 1), a bias slightly below $V_{c\downarrow}$.  
The condition for resonant tunneling is still met, conduction 
is allowed and the well is initially 
charged. A THz field of the same amplitude as for Fig. 1 is 
turned on at $t=0$, and as the radiation passes through the system 
the well empties, doing so in a few picoseconds time. 
Two characteristic times are involved in the data: the external  
radiation period $T_r=1/\nu$ and the time $T_w\sim$ 3 ps  it would take for our well to empty if 
charge is initially in it. When $T_r\gg T_w $ the radiation field essentially 
acts as an added dc bias and the well empties within the time $T_w$, while in the other  
extreme $T_r\ll T_w $ the oscillation is so fast that the electrons cannot respond and 
the system remains charged and conducting.

Figure 3 shows a situation in which the well is initially empty 
at a bias of $V$ = 0.285 V,  
slightly above the critical value $V_{c\uparrow}$ (empty square 
in Fig. 1). It is physically reached by lowering the bias after it has gone beyond 
$V_{c\downarrow}$. Once again the THz field is switched on at $t=0$.  
We observe that in all cases exhibited the well begins to charge, and after a 
transient time the system enters full resonance and current flows.

The above results assumed a radiation field of fixed amplitude $\delta E$ = 10 meV. 
One may ask how close to the critical value $V_{c\downarrow}$ must 
the system be biased in order to act as a switch for weaker radiation fields. 
This is shown in Fig. 4 for $\nu$ = 0.33 THz and assuming the switching to 
take place at $\tau$ = 17 ps  time. The bias offset is defined as 
$\delta V= V_{c\downarrow} - V$. The region above the curve
(labeled YES) is where the potential drop takes place within the time $\tau$,
while the region below (labeled NO) is where the switching does not
take place in that time interval. Within the range of our calculations
we found the shape of the curve in Fig. 4 to be generic,
shifting upwards as the frecuency increases. Close to the origin the 
dependence is approximately linear and for the chosen frequency
follows the relation 
$\delta E \sim \frac{1}{2} (1+\delta V)$. Using this expression we get that 
at a bias $\delta V$ = 1 meV our device would switch under radiation of about 
$50$ Watt/cm$^2$ and stronger. The sensitivity could be improved using 
 a wider collector barrier, thus having a narrower resonance 
(longer lifetime $T_w $). Because of limitations due to numerical instabilities 
this ansatz would be best tested experimentally. 
 
In summary, we have shown that an asymmetric double barrier heterostructure 
may act as a switch triggered by the passage of electromagnetic radiation at 
frequencies in the terahertz region and below. The frequency threshold for this  
switching action depends on the barrier and enclosed well widths. Depending 
on the applied external bias, the passage of current is turned on or off
by the radiation pulse. Our results rely on the current drop
as the resonance in the well falls below the emitter conduction band
edge, a feature also present in symmetric double barrier
heterostructures. In the latter case however, the drop is not an instability
of the system and does not take place abruptly, a desirable feature
for a switching device.
 
Work supported by FONDECYT grants No. 1990425 and No. 1990443.

\begin{figure}[h]
\caption{
 Current-Voltage characteristic for $\delta E$ = 0 (solid line)
 and $\delta E$ =10 meV(dashed line) at $\nu$ = 1 THz.
}
\end{figure}

\begin{figure}[h]
\caption{ 
 Time evolution of the charge density in the center of the 
well at $V$ = 0.310V
and frequencies 0.10 THz, 0.33 THz , 0.50 THz and 0.66THz.
}
\end{figure}

\begin{figure}[h]
\caption{ 
 Time evolution of the charge density in the center of the 
well at $V$ = 0.285 V
and frequencies 0.10 THz, 0.33 THz , 0.50 THz and 0.66 THz.
}
\end{figure}

\begin{figure}[h]
\caption{
 Boundary separating the region at which switching takes
place within 17 ps of the arrival of an external pulse at
frecuency $\nu$ = 0.33 THz, from the region in which the well
remains charged beyond that time interval.
}
\end{figure}
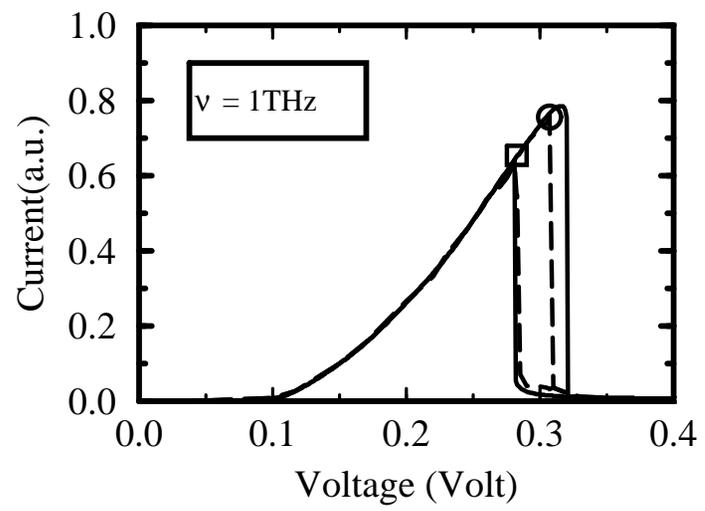
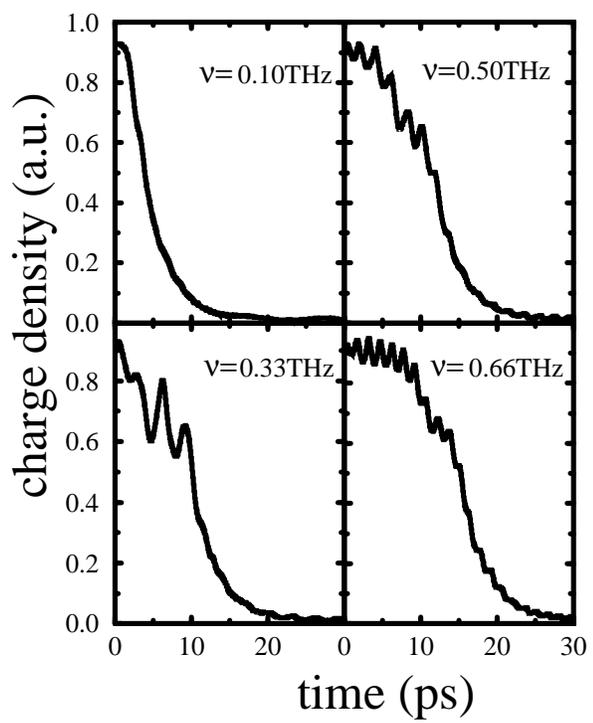
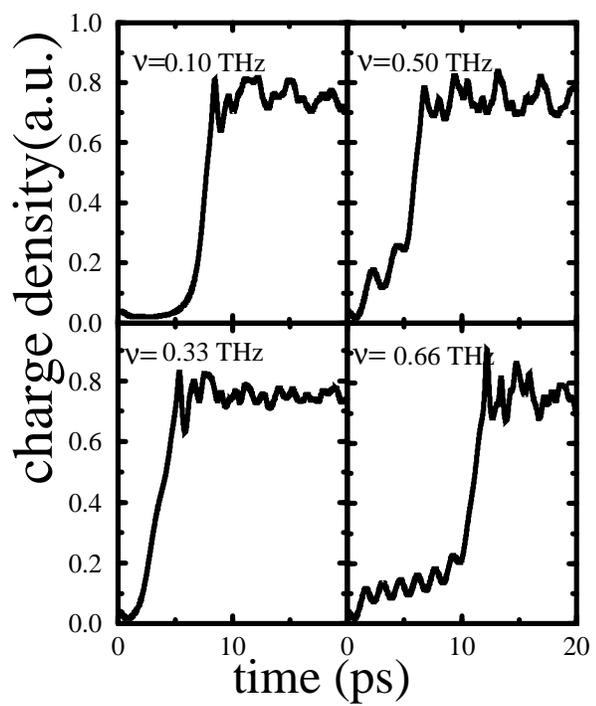
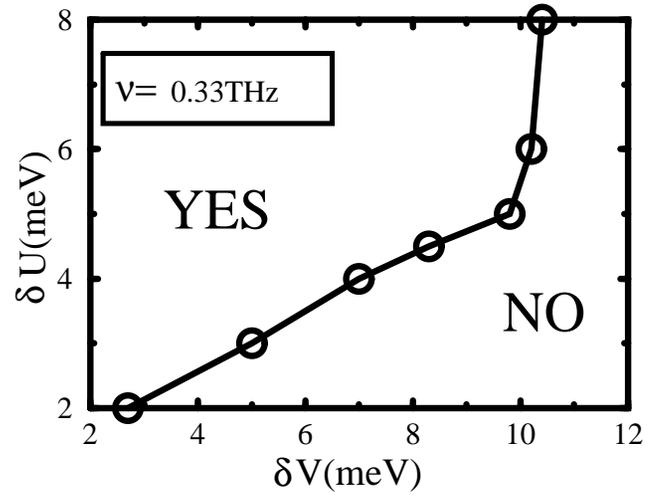
\end{document}